# Low-cost Stochastic Number Generators for Stochastic Computing

Sayed Ahmad Salehi, *Member, IEEE*
Electrical and Computer Engineering Department, University of Kentucky
Lexington, KY, USA

*Abstract—* Stochastic unary computing provides low-area circuits. However, the required area consuming stochastic number generators (SNGs) in these circuits can diminish their overall gain in area, particularly if several SNGs are required. We propose area-efficient SNGs by sharing the permuted output of one linear feedback shift register (LFSR) among several SNGs. With no hardware overhead, the proposed architecture generates stochastic bit streams with minimum stochastic computing correlation (SCC). Compared to the circular shifting approach presented in prior work, our approach produces stochastic bit streams with 67% less average SCC when a 10-bit LFSR is shared between two SNGs. To generalize our approach, we propose an algorithm to find a set of $m$ permutations ($n > m > 2$) with minimum pairwise SCC, for an $n$-bit LFSR. The search space for finding permutations with exact minimum SCC grows rapidly when $n$ increases and it is intractable to perform a search algorithm using accurately calculated pairwise SCC values, for $n > 9$. We propose a similarity function that can be used in the proposed search algorithm to quickly find a set of permutations with SCC values close to the minimum one. We evaluated our approach for several applications. The results show that, compared to prior work, it achieves lower MSE with the same (or even lower) area. Additionally, based on simulation results, we show that replacing the comparator component of an SNG circuit with a weighted binary generator can reduce SCC.

## I. INTRODUCTION

STOCHASTIC computing (SC) has emerged as an unconventional technique for performing computations by logic circuits [1]. Rather than performing computation on deterministic binary numbers, SC circuits are designed to process random bit streams. The input and output are represented by bit streams and their values are encoded as the probabilities of seeing 1's in the bit streams. Evidently, the values are confined in the unit interval [0,1], since probabilities cannot be beyond the unit interval. Compared to deterministic binary computing, SC provides several advantages including reduced hardware complexity and fault-tolerant computing. Because of these advantages, SC has been considered as an appropriate alternative to binary computing in different applications such as low-density parity check (LDPC) decoding [2], image processing [3], neural networks [4,5], and digital filters [6,7].

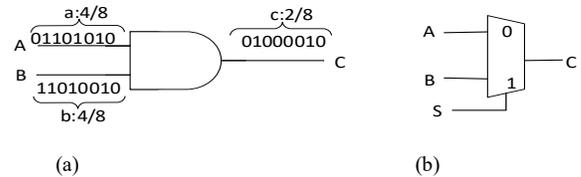

Fig.1 Low hardware-complexity in SC: (a) a simple AND gate computes multiplication, (b) a multiplexer compute scaled addition.

One main advantage of SC is its very low hardware-complexity that could result in cost-efficient computing circuits. The most common way to demonstrate the low hardware-cost of SC is its implementation of basic operations, i.e., multiplication and addition. Fig. 1(a) shows a simple AND gate implementing multiplication in SC. For the AND gate, the output is 1 only when input $A$ and input $B$ are both 1. Therefore, the probability of having 1 in the output bit stream is the multiplication of the probabilities of having 1 in each of the input bit streams, i.e., $P(C = 1) = P(A = 1) \times P(B = 1)$, that is $c = a \times b$. Similarly, Fig. 1(b) shows a 2-input multiplexer computing scaled addition. For the multiplexer, the output $C$ is 1 when $S$ is 0 and $A$ is 1 or when $S$ is 1 and $B$ is 1. Therefore, $P(C = 1) = (1 - P(S = 1)) \times P(A = 1) + P(S = 1) \times P(B = 1)$, that is $c = (1 - s) \times a + s \times b$.

A stochastic number generator (SNG) is an essential part of any SC circuit. A SC circuit uses SNGs to convert binary numbers to their corresponding random bit streams. They generate random bit streams with probabilities of producing 1's equal to their corresponding binary numbers. SNGs play a central role in the efficiency of a SC circuit for two reasons. First, for SC circuits the size of an SNG part is remarkable with respect to the computing part. This problem becomes more critical for applications with SC circuits that require many SNGs, such as high degree digital filtering and image processing algorithms. In fact, for several SC designs, SNG circuits consume around 80% or even 90% of the total area [8,9]. Second, the quality of the random numbers generated by SNGs can significantly affect the computational accuracy of SC designs and correlation among random bit streams is a source of inaccuracy in SC. Therefore, obtaining area-efficient and low-correlated SNGs is a major design challenge for SC.

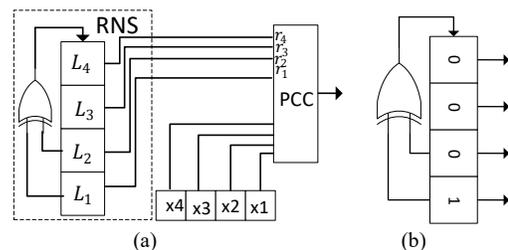

Fig. 2(a). General structure of a SNG. (b) LFSR is used as RNS.

In response to this challenge, the contributions of this paper are the following:

- Introducing a new permutation-based design space for sharing a random number source among several SNGs. The design space yields low-cost and low-correlated SNGs. Compared to SNGs with the same hardware complexity, the proposed SNGs generate random bit streams with lower cross correlation.

- Modeling the variation of SC correlation for the proposed design space and presenting a searching algorithm for finding the permutations with minimum correlation. In addition, we present a similarity function that can be used to speed up the searching algorithm by degrading its accuracy in obtaining the permutations with exact minimum SC correlation. Even the fast version of the proposed algorithm achieves permutations with lower SC correlation, compared to prior work with the same hardware complexity.

- Using simulation results to demonstrate a reduction in SC correlation achievable by replacing the comparator component of SNG circuits with weighted binary generator.

In the next section, we explain the general structure of SNGs, a measure to evaluate their performance in SC, and related prior work. Section III describes the proposed design technique for two SNGs sharing a random number source. Section IV presents a low computational complexity model for the correlation variation of the proposed design and Section V generalizes the design approach for more than two SNGs. In Section VI, we evaluate our technique for some applications and Section VII concludes the paper.

## II. PRELIMINARIES AND PRIOR WORK

### A. SNG

Generally speaking, an SNG is composed of two parts: a random number source (RNS) and a probability conversion circuit (PCC). An RNS is used to generate a sequence of uniformly distributed random numbers, while a PCC is designed to convert the generated random numbers into a random bit stream with the desired probability of generating 1's. Fig. 2(a) shows an SNG circuit.

A linear feedback shift register (LFSR) and a cellular automata (CA) can be used as a digital RNS. A CA is made up of cascaded modules called a cell or site [10]. Each cell is composed of a flip-flop and a combinational circuit. In its simplest form, each cell is connected to only two neighbor cells on its left and right. The next value of one cell is defined by its current value and that of the connected neighbor cells. Although a CA provides modularity and can generate good-quality random numbers, it is not commonly used in SC circuits due to its hardware-complexity compared to an LFSR. Due to the low hardware-complexity and high speed of an LFSR, it is employed as the RNS part in most SC circuits, including the circuits proposed in this paper. The advantage of using an LFSR is more crucial for computationally intensive applications such as deep neural networks [5] and energy-limited applications such as embedded systems and mobile internet of things (IoT) devices. Note that CA and LFSR circuits cannot be designed to generate true random numbers; however, their output sequences pass some of the random number tests and if the period of the sequences is large enough, they resemble an ideal RNS for SC computing [11].

An $n$-bit LFSR is composed of an $n$-bit shift register and one or more XOR gates. Fig. 2 (b) shows a 4-bit LFSR initialized by 0001. Normally, an LFSR is designed to have the maximum sequence length of $2^n - 1$. That is, the output sequence of a maximal-length LFSR repeats after a period of $2^n - 1$ binary numbers and each number in the range of $[1, 2^n - 1]$ is generated once in the period.

Considering the sequence of bits produced in each single flip-flop of the LFSR in Fig. 2 (b), four random bit streams are generated as shown in Table I by $L_4, L_3, L_2,$ and $L_1$. Each bit stream has a period of 15 bits including eight 1's and seven 0's.

In general, for an $n$-bit maximal-length LFSR, the bit pattern in each bit stream repeats every $2^n - 1$ bits. Since $2^{n-1}$ of bits in the pattern are 1's, the probability of nearly 0.5 is generated by each bit stream.

Because all the bit streams produced by an LFSR have the probability of 0.5, a PCC is required in order to generate a bit stream with a desired probability other than a 0.5. PCC is a combinational circuit with two $n$-bit inputs. One is connected to a deterministic binary number $x$, and the other one to a sequence of numbers generated by an LFSR. It produces a bit stream with the probability of $P_x = x \times 2^{-n}$, or more accurately $P_x = \frac{x}{2^n - 1}$. In each clock cycle, one $n$-bit number from the output sequence of an LFSR is converted to one bit. The output bit stream is generated such that the total of $x$ bits in each period are 1 and the other bits are 0. In the literature of SC, two types of PCC circuits have been proposed: digital comparator (CMP) and weighted binary generator (WBG). A CMP is an $n$-bit digital comparator circuit that produces a 1 if the random number from the LFSR is less than the binary number $x$, and a 0 otherwise. A WBG circuit works differently. First it converts the output sequence of an LFSR into a sequence of weighted binary numbers and then, generates the output bit stream using the weighted sequence and input binary number $x$ [12]. For a 4-bit CMP and WBG, the internal circuits are illustrated in Fig. 3(a) and 3(b), respectively. Although both circuits generate bit streams with the desired probability for every input $x$, their internal logic circuits and generated bit streams are different. For $x = 1011$, Table I tabulates the output bits generated by a CMP, $S_{CMP}$, and a WBG, $S_{WBG}$, for an LFSR's output, $L_4 L_3 L_2 L_1$. In this paper, we examine both CMP and WBG circuits as the PCC part of the proposed SNG circuits.

TABLE I
THE OUTPUT BITSTREAMS FOR THE LFSR IN FIG. 2 WHEN A COMPARATOR OR A WBG ARE USED AS THE PCC PART.

| L4 | L3 | L2 | L1 | $S_{CMP}$ | $S_{WBG}$ |
|---|---|---|---|---|---|
| 0 | 0 | 0 | 1 | 1 | 1 |
| 1 | 0 | 0 | 0 | 1 | 1 |
| 0 | 1 | 0 | 0 | 1 | 0 |
| 0 | 0 | 1 | 0 | 1 | 1 |
| 1 | 0 | 0 | 1 | 1 | 1 |
| 1 | 1 | 0 | 0 | 0 | 1 |
| 0 | 1 | 1 | 0 | 1 | 0 |
| 1 | 0 | 1 | 1 | 1 | 1 |
| 0 | 1 | 0 | 1 | 1 | 0 |
| 1 | 0 | 1 | 0 | 1 | 1 |
| 1 | 1 | 0 | 1 | 0 | 1 |
| 1 | 1 | 1 | 0 | 0 | 1 |
| 1 | 1 | 1 | 1 | 0 | 1 |
| 0 | 1 | 1 | 1 | 1 | 0 |
| 0 | 0 | 1 | 1 | 1 | 1 |
| 8/15 | 8/15 | 8/15 | 8/15 | 11/15 | 11/15 |

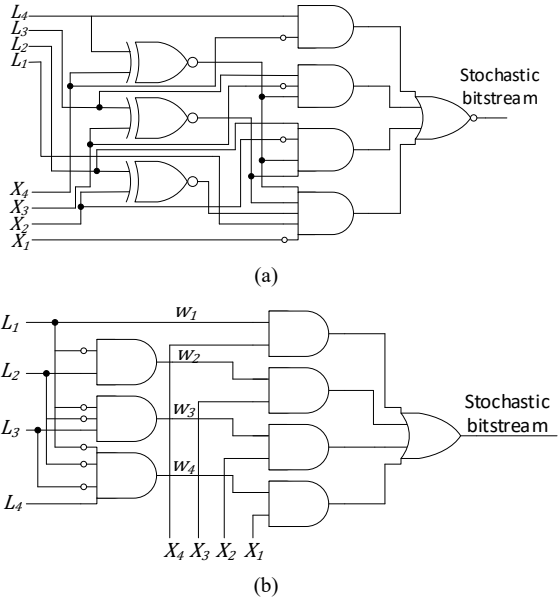

Fig. 3 Two commonly used PCC circuits: (a) CMP, (b) WBG.

### B. SCC

When two (or more) random bit streams are used as inputs for a SC circuit, the cross correlation between them can affect the computational accuracy of the circuit. Assume $s_x$ is a random bit stream generated for binary number $x$ and $s_y$ is generated for binary number $y$. In order to quantitatively evaluate the correlation between $s_x$ and $s_y$, one commonly used measure is the SC correlation (SCC) computed by (1) [13].

Where, $P(s_x)$ and $P(s_y)$ are, respectively, the probabilities for bit streams $s_x$ and $s_y$ to have 1's and $\delta(S_x, S_y) = P(S_x \wedge S_y) - P(s_x)P(s_y)$ with $\wedge$ denotes the bitwise AND of $s_x$ and $s_y$. SCC can have values between -1 and +1, where $\pm 1$ indicate maximum correlation and 0 means no correlation. When comparing the corresponding bits of the two bit streams, the SCC is positive if most 1's and 0's are aligned. However, if

$$SCC(S_x, S_y) = \begin{cases} \frac{\delta(S_x, S_y)}{\min(P(s_x), P(s_y)) - P(s_x)P(s_y)} & \delta(S_x, S_y) => 0 \\ \frac{\delta(S_x, S_y)}{P(s_x)P(s_y) - \max(P(s_x) + P(s_y) - 1, 0)} & \delta(S_x, S_y) < 0 \end{cases} \quad (1)$$

most corresponding bits are complemented to each other, the SCC is negative. Since lower absolute SCC values elicit more accurate results in SC, researchers seek designs that generate bit streams with low SCCs. In general, the absolute values for SCC among bit streams generated in each flip-flop of an LFSR (before connecting them to a PCC) are low. For example, the cross correlation between each pair of bit streams generated by a maximal-length 4-bit LFSR, e.g., $(L_2, L_1)$ in Table I, is -0.0816 and becomes smaller as $n$ increases.

As suggested in [6], we evaluate the correlation between two SNGs by finding the average SCC among their generated bit streams for all possible input values and represent it as $SCC_{avg}$. We can calculate the $SCC_{avg}$ for SNG1 and SNG2 by (2).

$$SCC_{avg}(SNG1, SNG2) = \sum_{i=0}^{2^n-1} \sum_{j=0}^{2^n-1} \frac{|SCC(s_i, s'_j)|}{2^n \times 2^n} \quad (2)$$

Where, $s_i$ and $s'_j$ are bit streams generated by SNG1 and SNG2, respectively. To calculate the $SCC_{avg}$, first, for both SNGs, the bit streams of all possible inputs, i.e., $s_k$ and $s'_k$ for $k = 1, 2, \ldots, 2^n - 1$, are generated. Then, SCC values between each bit stream of SNG1 and bit streams of SNG2 are calculated by (1). Finally, the $SCC_{avg}$ is calculated by computing and normalizing the total sum of the SCCs. Obviously, $SCC_{avg}$ is a positive number between 0 and 1 and the lower its value means less correlation between the two SNGs.

### C. Prior Work

When several bit streams are required in a SC circuit, the straightforward implementation is to use a separate SNG to generate each bit stream. [14] has shown that careful seeding, scrambling, and feedback polynomials for the LFSR parts of these SNGs can improve computational accuracy. However, rather than using a separate LFSR for each SNG, a common approach to design compact SNGs is based on sharing an LFSR among them. Although sharing an LFSR reduces the hardware cost, it significantly raises the cross correlation between each pair of the generated random bit streams and thus leads to computational inaccuracy. It is worth mentioning there are a limited number of applications for SC where computational accuracy is not affected by the correlation between bit streams. Therefore, an LFSR can be directly shared among different SNGs [15]. However, it is required for many applications to reduce the mutual correlation among random number sequences before sharing them [5]. [16] has suggested using an extra S-Box circuit to generate low-correlated copies of an LFSR's output to be shared with different SNGs. Although this method generates low auto- and cross-correlated bit streams, the S-Box is a combinational circuit that can increase the hardware complexity significantly for large values of $n$. Recent work [8] has suggested a circular shifting approach in order to obtain bit streams with low cross correlation from a shared LFSR without hardware overhead. However, the approach does not provide bit streams with the minimum cross correlation. In fact, circular shifts are a small portion of an unexplored design space that can produce low-correlated bit streams from a shared LFSR with no hardware overhead. This research investigates the whole design space to find designs with minimum $SCC_{avg}$.

Note that some SNGs [17] generate multi-bit-width (parallel) bit streams for an input binary number, but this paper focuses on SNGs generating single-bit-width (serial) bit stream.

### III. SNG COST REDUCTION WITH PERMUTATION-BASED SHARED RNS

This section presents the proposed approach for the design of efficient SNGs. We share one LFSR between two SNGs to reduce the area cost. However, unlike prior work, we reduce the correlation among the generated sequences without adding any extra hardware. In this section, we explain the method for two SNGs and in Section V, we generalize the idea for designing more than two SNGs.

In order to generate low-correlated random bit streams, the cross correlation among the sequences of random numbers fed to the PCCs of different SNGs should be low. While using one LFSR generates one sequence of random numbers, we can feed different sequences of random numbers to different PCCs by permuting the connection between the LFSR's output and the inputs of the PCCs. Consider a simple example of generating two random bit streams from a 4-bit LFSR. Fig. 4(a) shows two 4-bit SNGs sharing one LFSR based on the proposed approach. We connect the $L_4, L_3, L_2,$ and $L_1$ outputs of the LFSR, to the $r_4, r_3, r_2,$ and $r_1$ inputs of the first PCC, respectively. For the second SNG, however, we permute the output bits of the LFSR before connecting them to the inputs of the SNG's PCC.

The permutations of an LFSR's output can provide low-correlated random number sequences with the required feature due to two reasons. First, in general, there is a low correlation among bits in the flip-flops of an LFSR at a given time. Thus, permuted versions of LFSR's output sequences have low cross correlation and they feed low-correlated sequences of random numbers to different PCCs. Second, all permutations of a maximal-length LFSR generate uniformly distributed random numbers such that in its repeating cycle (period), every integer binary number between 1 and $2^n - 1$ is repeated exactly once. Hence, the permutation of an LFSR's output bits does not affect the functionality of SNGs connected to them. That is, in each period, connected PCCs generate the desired number of 1's and 0's but in a permuted order.

The approach can be extended to any $n$-bit maximal-length LFSR: the first SNG is built by direct connection of the LFSR's output to a PCC's input whereas the second SNG is built by connecting the permuted output of the LFSR to another PCC's input. The permuted output should be chosen such that the SCC between the generated bit streams of the first and second SNG is minimum. However, other than the direct connection, there are $(n! - 1)$ different permutations for an $n$-bit LFSR output; which one achieves the minimum $SCC_{avg}$? To answer this question for different values of $n$, we examine all possible permuted connections of the LFSR and search for those with minimum $SCC_{avg}$. We start for the case of $n = 4$. Assume that vector $L = [L_1, L_2, L_3, L_4]$ is the output of a 4-bit LFSR. There are 24 possible permutations for L. Also, assume the first SNG is formed by connecting $L_4, L_3, L_2,$ and $L_1$, respectively, to $r_4, r_3, r_2,$ and $r_1$ of a PCC. Among the other 23 possible permutations for L, the SNG that results in the minimum $SCC_{avg}$ with the first SNG, is formed by connecting $L_1, L_2, L_3,$ and $L_4$, respectively, to $r_4, r_3, r_2,$ and $r_1$, of another PCC. Similar results are observed by investigating the proposed approach for the permutation of other values of $n$. The following conclusion generalizes the approach: for $i = 1, 2, ..., n$, if the first SNG is formed by connecting $L_i$, i.e., the $i$th flip-flop of an LFSR, to $r_i$, i.e., the $i$th input of a PCC, then the second SNG, resulting in the minimum $SCC_{avg}$ with the first SNG, is formed by connecting $L_{n-i}$ output of the LFSR to $r_i$ input of another PCC. [18] has proved that a permutation with reversed ordering provides the maximum deviation distance that agrees with our results.

For example, to share an 8-bit LFSR, $L_1$ to $L_8$ are respectively connected to the $r_1$ to $r_8$ of a PCC to build the first SNG and $L_8$ to $L_1$ are respectively connected to the $r_1$ to $r_8$ of another PCC to build the second SNG. Fig. 4 shows the proposed LFSR-sharing approach based on permutation for $n=4$ and $n=8$.

To illustrate how the $SCC_{avg}$ varies with permutation, Fig. 5 (a)-(d) shows the $SCC_{avg}$ values between the first SNG and the permuted ones for $n=4$ to 7. For the purpose of better readability, we do not include the graph of $SCC_{avg}$ for higher values of $n$; however, they have a similar pattern. The horizontal axis ranges from 1 to $n!$ and represents the index of permutation in reverse lexicographic order [19] (the same order produced by MATLAB's function *perms*) [20]. For reverse lexicographic order, the permutation of a vector is performed based on the positional index of its elements. That is, the permuted versions of a vector are formed by rearranging its elements from left to right and starting from greater positional indices. For example, for $n=4$ and original vector [1, 2, 3, 4], the permutations in reverse lexicographic order are listed as: [4,3,2,1], [4,3,1,2], [4,2,3,1], [4,2,1,3], [4,1,3,2], [4,1,2,3], [3,4,2,1], [3,4,1,2], [3,2,4,1], [3,2,1,4], [3,1,4,2], [3,1,2,4], [2,4,3,1], [2,4,1,3], [2,3,4,1], [2,3,1,4], [2,1,4,3], [2,1,3,4], [1,4,3,2], [1,4,2,3], [1,3,4,2], [1,3,2,4], [1,2,4,3], [1,2,3,4].

So, the first $SCC_{avg}$ is corresponding to the permuted vector [4, 3, 2, 1], the second one to the permuted vector [4, 3, 1, 2], and so on. Note that in the reverse lexicographic order, the last permuted vector is the same as the original vector.

For all values of $n$ in Fig. 5, the $SCC_{avg}$ corresponding to the first permutation is the minimum $SCC_{avg}$. This permutation is representing the connection of $L_n, L_{n-1}, \ldots, L_1$ to $r_1, r_2, \ldots, r_n$. On the other side, since the last permutation, i.e., vector $[1, 2, \ldots, n]$, is the same as the original SNG connection, it has the maximum $SCC_{avg}$.

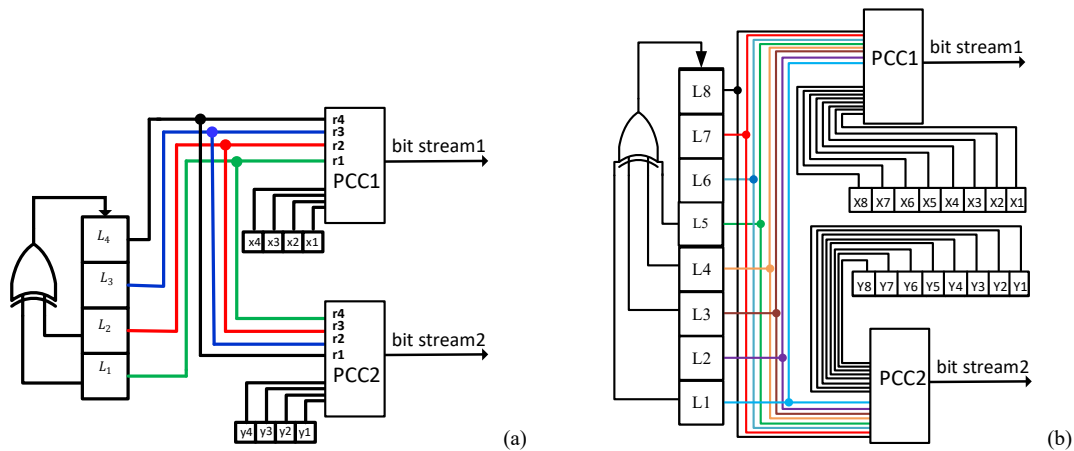

Fig. 4- Proposed structure for sharing an LFSR with two SNGs based on output permutation: (a) $n$=4 and (b) $n$=8.

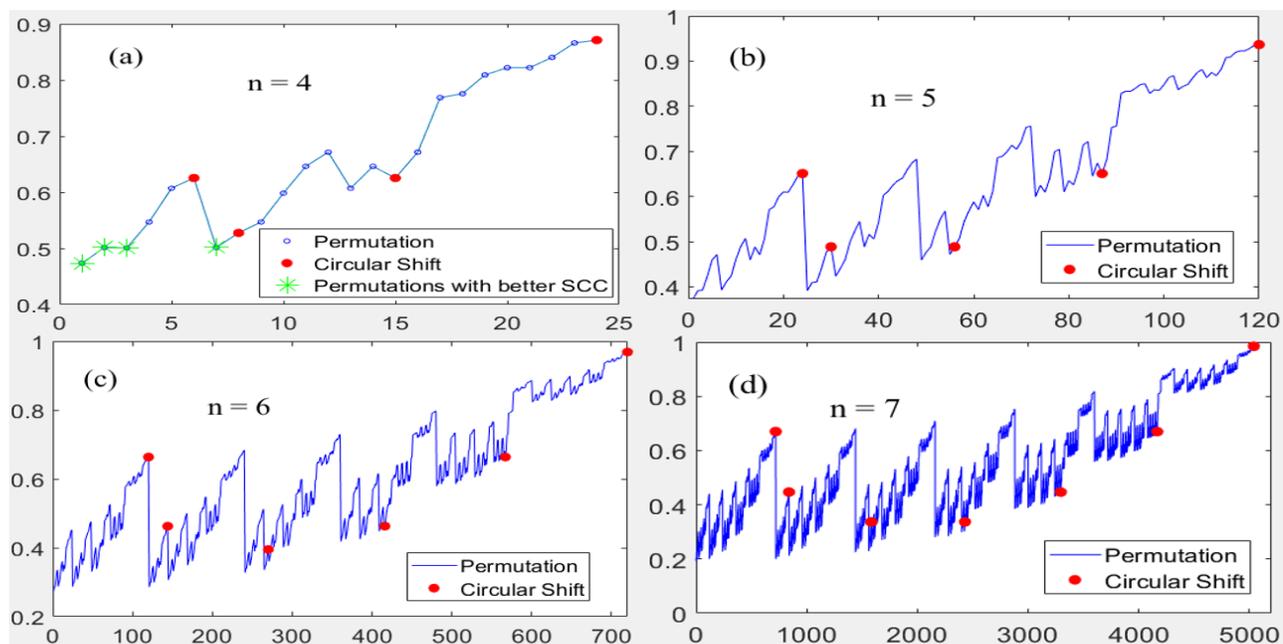

Fig.5 The $SCC_{avg}$ variation for circular shift and permutation-based methods.

Let $k$, where $1 \leq k \leq n$, denote the number of shifts in the circular shifting approach [8]. The red dots in Fig. 5 mark the values of $SCC_{avg}$ related to the circular shifts with $k$ bits shift. As it is explained in [8], compared to the other values of $k$, the circular shift with maximum gap, i.e., $k = n/2$, yields the lowest $SCC_{avg}$ values achievable by the circular shifting approach. Yet, our proposed permutation-based approach can find $SCC_{avg}$ values lower than those produced by the circular shifting approach. The green stars in Fig. 5(a) mark these points for $n$=4. The minimum $SCC_{avg}$ calculated for $n$=4 to $n$=10 is listed in Table II. The first and second columns compare the minimum $SCC_{avg}$ achievable by the circular shift and our permutation approaches. For both methods an $n$-bit comparator is used as the PCC part. The third column is for using an $n$-bit WBG as the PCC part in our method. Obviously, using WBG as the PCC part and increasing the value of $n$ further reduce the $SCC_{avg}$.

IV. MODELING AND ANALYSIS

To find the permutation with minimum $SCC_{avg}$, we need to calculate $SCC_{avg}$ between the original SNG and $(n! - 1)$ other SNGs formed from the permutation of an $n$-bit LFSR's output. As the length of LFSR increases, the search space and the required resources to find the solution rapidly become much larger. For example, for $n = 11$, each copy of all permutations of L requires more than 3 GB of RAM [20] and it grows quickly. In fact, for $n > 9$, an exhaustive search for finding the minimum $SCC_{avg}$ is intractable. In order to reduce the computational complexity of this problem, we model the behavior of $SCC_{avg}$ for different permutations of an LFSR by introducing a new function; one that we call the *similarity function*. Assuming the original (non-permuted) output vector of an $n$-bit LFSR is $L=[L_1, L_2, \ldots, L_n]$, the positional index for $L_1$ is 1, for $L_2$ is 2, and so on. Also, let $PL_k$, for $1 \leq k \leq n!$, denote the $k$th permuted vector of L in the reverse lexicographic order. The similarity function calculates an approximation of the $SCC_{avg}$ between L and its permutations, $PL_k$, and is defined as:

$$S(k) = \sum_{i=1}^{n} i \times ind(PL_k(i)) \quad (3)$$

where, $ind(PL_k(i))$ is the index of the $i$th element of vector $PL_k$ in the original vector L. For example, if $k = 1$ then $S(1)$ calculates the similarity function for $PL_1$, the first permutation of L. Since $PL_1=[L_n, L_{n-1}, \ldots, L_1]$, the $i$th element of $PL_1$ is $L_{(n-i+1)}$. That is to say, the index for the $i$th element of $PL_1$ is $(n - i + 1)$ in the original vector L. Therefore, $S(1)$ is calculated as $S(1) = \sum_{i=1}^{n} i \times (n - i + 1)$. Similarly, if $k=2$, $S(2)$ is calculated as $S(2) = (n - 1) + 2n + \sum_{i=1}^{n-2} i \times (n - i + 1)$, because $PL_2 = [L_n, L_{n-1}, \ldots, L_3, L_1, L_2]$. For the last permutation of L, i.e., $PL_{(n!)}$, since it is the same as the original vector L, $S(n!)$ is calculated as $S(n!) = \sum_{i=1}^{n} i \times i$.

The value of $S(k)$ is smaller if more elements of the corresponding permuted vector, $PL_k$, change their positional index with respect to the original vector L. In other words, the similarity function is smaller if, in the connection of LFSR's output to PCC unit, more bits are permuted with respect to the direct connection. Therefore, the similarity function provides an estimate of the correlation among bit streams generated by the permutations of an LFSR. Fig. 6 illustrates the graph of normalized $S(k)$ for $n$=4 to 7. The figure includes the graph of the $SCC_{avg}$ values to make comparison easier. Although the similarity function does not calculate the exact values for the $SCC_{avg}$, comparing two graphs shows that this function approximately models the behavior of the $SCC_{avg}$ and provides an approximation of indices of permutations with the minimum $SCC_{avg}$. We can find other measures for the closeness between permutations of a sequence [21]. Among them, squared deviation distance [22] achieves the same pattern as $S(k)$, but with different calculations, and can be used for our model. Compared to the other measures, $S(k)$ appropriately approximates the $SCC_{avg}$ with lower computational complexity.

In fact, the similarity function forms a low-cost heuristic computation approach for the estimation of the $SCC_{avg}$. For any reason, such as limitations in circuit level implementation, if a designer decides to use permutations other than the permutation with the minimum correlation, then the similarity function can provide a guiding estimate for choosing other permutations with low correlation.

TABLE II
COMPARISON OF MINIMUM $SCC_{avg}$_AVG VALUES ACHIEVABLE BY CIRCULAR SHIFT [8] AND THE PROPOSED APPROACH.

| $n$ | Circular Shift [8] | Proposed (CMP) | Proposed (WBG) |
|---|---|---|---|
| 4 | 0.528 | 0.473 | 0.387 |
| 5 | 0.467 | 0.372 | 0.286 |
| 6 | 0.387 | 0.274 | 0.198 |
| 7 | 0.336 | 0.192 | 0.132 |
| 8 | 0.270 | 0.130 | 0.085 |
| 9 | 0.218 | 0.086 | 0.053 |
| 10 | 0.162 | 0.054 | 0.033 |

Notice that for any specific $n$, as long as each number between 1 and $2^n - 1$ repeats exactly once in each period of the random number sequence, the variation of $SCC_{avg}$ with respect to permutations is the same. That is, for all possible structures of an $n$-bit LFSR, if it is a maximal-length LFSR, the values of the $SCC_{avg}$ for permutations are the same. Further, the behavior of the $SCC_{avg}$ is independent of which permutation we choose as the original (direct) output of the LFSR. For example, if we choose L=$[L_2, L_1, L_4, L_3]$ as the direct output of a 4-bit LFSR, then the permutation $[L_3, L_4, L_1, L_2]$ provides the minimum $SCC_{avg}$ with L.

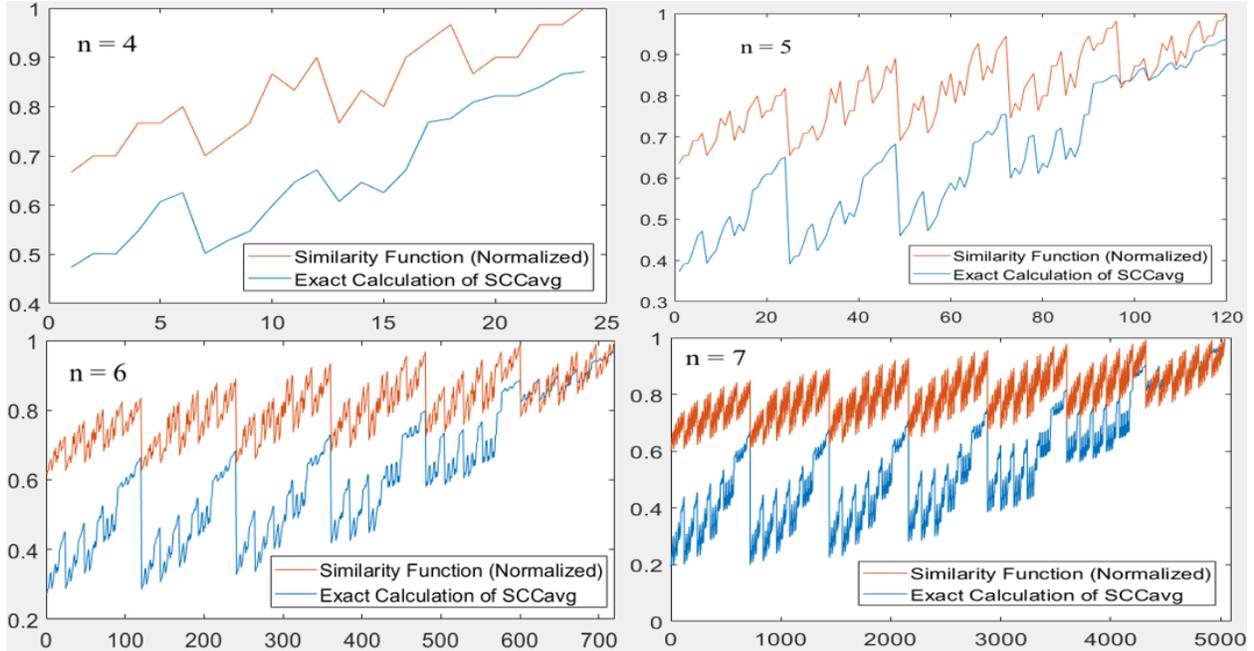

Fig.6 -The exact $SCC_{avg}$ values and the similarity function for different values of $n$.

## V. GENERALIZATION

So far, we have discussed the permutation-based sharing of an LFSR between two SNGs. However, the idea can be extended to sharing an $n$-bit LFSR for more than two SNGs. Let us assume we want to share an LFSR among $m$ SNGs, where $n > m > 2$. The goal is to find a set of $m$ different permutations of the LFSR's output such that the maximum of all pairwise $SCC_{avg}$ values for this set is minimum among other possible sets. For example, assume $m = 3$ and $P1, P2$, and $P3$ are indices in reverse lexicographic order for the permutations of an $n$-bit LFSR that build 3 SNGs with the minimum mutual values of $SCC_{avg}$. Then, permutations $PL_{P1}$, $PL_{P2}$, and $PL_{P3}$ are the ones that minimize $MP_3(PL_{Pi}, PL_{Pj}, PL_{Pk})$, where
$MP_3(PL_{Pi}, PL_{Pj}, PL_{Pk}) =$
$\max\{SCC_{avg}(PL_{Pi}, PL_{Pj}), SCC_{avg}(PL_{Pi}, PL_{Pk}), SCC_{avg}(PL_{Pj}, PL_{Pk})\}$
for $1 \leq i, j, k \leq n!$.

The pseudo code in Algorithm 1 represents the proposed algorithm for finding a set of $m$ permutations that can provide a RNS for $m$ SNGs with minimum $SCC_{avg}$ values. For each permutation, the algorithm examines whether it can be part of a set of $m$ permutations with minimum $SCC_{avg}$ values. It starts with SM=1, the greatest possible value for $SCC_{avg}$. If the maximum value of $SCC_{avg}$ among a set of $m$ permutations is less than SM, then the algorithm updates SM with this maximum value and saves the indices of the permutations in $P1, P2, \ldots, Pm$ as the current set with the minimum $SCC_{avg}$ value. This process repeats for all possible sets of permutations and after examining the last set, indices for the best set are saved in $P1, P2, \ldots, Pm$. As an example, Algorithm 2 represents the pseudo code for $m = 3$. The pairwise $SCC_{avg}$ values and indices of three permutations with minimum mutual $SCC_{avg}$ values for $n$=4 to 7 are listed in Table III. We extend the circular shift approach for obtaining a set of 3 shifts with minimum $SCC_{avg}$ values and list the results in Table IV. Comparing the results in Tables III and IV show that the proposed method can achieve triple sets with lower $SCC_{avg}$ values. Notice that for both methods there are more than one set with minimum $SCC_{avg}$ values. Running the algorithm and replacing CMP by WBG in the permutation-based SNGs reduces the obtained $SCC_{avg}$ values even more as listed in Table V.

Here, we assume the cross correlations between all elements in a set of $m$ permutations are equally important. However, if it is required for particular applications, we can change the

**Algorithm 1: Algorithm for finding the set of $m$ permutations of a LFSR with minimum pairwise $SCC_{avg}$.**

**Input:** $m$
**Outputs:** $P1, P2, \ldots, Pm$
**Initialization:** SM=1
for $i = m$ $to$ $n!$ do
  for $j = 1$ $to$ $i - (m - 1)$ do
    if $SCC_{avg}(PL_i, PL_j) <$ SM then
      $SP_1 = SCC_{avg}(PL_i, PL_j)$;
      for each set $(k_1, k_2, \ldots, k_{m-2})$ of $m - 2$
      elements in $\{j + 1, j + 2, \ldots, i - 1\}$ do
        if $SCC_{avg}(PL_i, PL_{k_r}) <$ SM for all $k_r \in (k_1, k_2, \ldots, k_{m-2})$ then
          for all $k_r$ do
            $SP_{k_r} = SCC_{avg}(PL_i, PL_{k_r})$;
            if $SCC_{avg}(PL_{h_1}, PL_{h_2}) <$SM for all pairs$(h_1, h_2)$
            in $(j, k_1, k_2, \ldots, k_{m-2})$ then
              $SP_{h_1 h_2} = SCC_{avg}(PL_{h_1}, PL_{h_2})$;
              $P1 = i$;
              $P2 = j$;
              $P3 = k_1$;
              $\vdots$
              $Pm = k_{m-2}$
              SM= $\max$ {all $SP_q$s: $SP_1, SP_{k_1}, \ldots, SP_{k_{m-2}}, SP_{h_1 h_2}, \ldots$}

**Algorithm 2: Algorithm for finding the set of 3 permutations of an LFSR with minimum pairwise $SCC_{avg}$.**

**Initialization:** SM=1
for $i = 3$ to $n!$ do
  for $j = 1$ to $i - 2$ do
    if $SCC_{avg}(PL_i, PL_j) <$SM then
      $SP_1 = SCC_{avg}(PL_i, PL_j)$;
      for $k_1 = 1$ to $i - j - 1$ do
        if $SCC_{avg}(PL_i, PL_{j+k_1}) <$ SM then
          $SP_{k_1} = SCC_{avg}(PL_i, PL_{j+k_1})$;
          if $SCC_{avg}(PL_j, PL_{j+k_1}) <$SM then
            $SP_{h_1 h_2} = SCC_{avg}(PL_j, PL_{j+k_1})$;
            $P1 = i$;
            $P2 = j$;
            $P3 = j + k_1$;
            SM= $\max$ { $SP_1, SP_{k_1}, SP_{h_1 h_2}$}

algorithm to give priority to $SCC_{avg}$ values for some pairs over others.

Due to the inequalities in the *if* statements of Algorithm 1, part of its pseudocode is executed conditionally. That is, different permutations may require different amounts of time to complete their pass in the algorithm. To estimate the worst-case time complexity of finding the best set of $m$ permutations, we break down the process into four steps:

1) Calculating all possible permutations for an $n$-bit LFSR; the computational complexity is $O(n! \times n)$. 2) Calculating bit streams for each permutation; the computational complexity is $O(2^{2n})$. 3) Calculating $SCCavg$ for all pairs of the permutations; the computational complexity is $O((n! \times 2^n)^2)$.

4) Finding the set of $m$ permutations with minimum pairwise $SCC_{avg}$ by performing comparisons. To calculate the number of required computations, assume $MP_m$ is the maximum of all pairwise $SCC_{avg}$ values for a set of $m$ permutations. Since there are $\binom{m}{2}$ distinct pairs of permutations within a set, finding $MP_m$ for each set requires $\binom{m}{2} - 1$ comparisons. Since, for an $n$-bit LFSR, $\binom{n!}{m}$ distinct sets of $m$ permutations are possible, the total number of required comparisons to find maximum values for all the sets is $\binom{n!}{m} \times [\binom{m}{2} - 1]$. Finally, finding the minimum of the maximum values requires $\binom{n!}{m} - 1$ comparisons. Thus, the total number of required comparisons by the algorithm is $\binom{n!}{m} \times [\binom{m}{2} - 1] + \binom{n!}{m} - 1$, simplified to $\binom{n!}{m} \times \binom{m}{2} - 1$.

The total computational complexity for the worst-case runtime is the sum of the above four steps. Note that when $n$ is increased, in addition to the required computational complexity, the required memory grows exponentially and becomes a challenge. To show the actual runtime of Algorithm 1, Table VI lists the runtime of the algorithm for different values of $n$ and $m$ implemented by MATLAB on a computer with a Core i7 2.11GHz intel processor and 16 GB of RAM.

Rather than using $SCC_{avg}$, we can use the similarity function in Algorithm 1 to reduce its runtime. We replace steps 1, 2 and 3 by calculating $S(k)$ using (3). First, we use $S(k)$ to find the indices of the best $m$ permutations and then compute the exact value of $SCC_{avg}$ for these indices. Table VI shows the average runtime using $S(k)$ for different values of $n$ and $m$. As the table shows, using $S(k)$ significantly reduces the computational time of the algorithm. The reduction becomes more significant when $n$ increases. Table VII shows the indices and values of the best three permutations obtained using $S(k)$ in Algorithm 1. These results show that by using the similarity function we can find permutations with $SCC_{avg}$ values very close to those listed in Table III. In fact, the $SCC_{avg}$ values of the achieved permutations for $n$=5 is the same in both Table III and Table VII. Although using the similarity function in Algorithm 1 does not necessarily achieve the minimum correlations, it achieves correlation values lower than circular shifting results listed in Table VI.

## VI. EVALUATION WITH APPLICATIONS

In this section, we evaluate the proposed design approach using applications with different levels of complexity ranging from simple multiplication to image segmentation. For all experiments, we used 8-bit maximal-length LFSRs and represent variables by 255-bit random bit streams. To make a fair comparison with prior work regarding hardware implementation, we use the synthesis results obtained by Synopsys Design Compiler in 45nm NanGate library [23]. We compare the results for 6 different methods: deterministic (conventional binary), no-share LFSR (separate LFSR for each SNG), simple-share (one LFSR with the same output connection for all SNGs), SBoNG [16], circular shift [8], and our proposed method. Fig. 7 shows the circuit area (in $\mu m^2$) and Table VIII lists the mean-squared error (MSE) for each application.

As the first application, we implement a simple 2-input multiplier. For the binary multiplication, we use a conventional $8 \times 8$ Wallace tree multiplier [24]. Because the MSE varies for some SC-based circuits due to the use of different LFSRs, we calculate the average MSE for 1000 trials with different LFSRs. While the proposed circuit has the same size as the simple-share and circular shift circuits, it achieves higher accuracy. Interestingly, the multiplier implemented using the proposed method has a lower MSE than the no-share and SBoNG methods. We can justify this by considering the fact that in our method we design SNGs based on minimum SCC values and the definition of SCC in (1) is completely in favor of multiplication of two bitstreams. Thus, we find a pair of an LFSR's permutations that can yield better results than two separate LFSRs or randomized output of an LFSR for multiplication.

TABLE III
MINIMUM $SCC_{avg}$ VALUES FOR SHARING AN LFSR WITH 3 DIFFERENT SNGS USING THE PROPOSED APPROACH WITH COMPARATOR.

| $n$ | P1 | P2 | P3 | $SCC_{avg}$ $(PL_{P1}, PL_{P2})$ | $SCC_{avg}$ $(PL_{P1}, PL_{P3})$ | $SCC_{avg}$ $(PL_{P2}, PL_{P3})$ |
|---|---|---|---|---|---|---|
| 4 | 4 | 9 | 24 | 0.5470 | 0.5470 | 0.5470 |
| 5 | 12 | 44 | 88 | 0.4887 | 0.4882 | 0.4885 |
| 6 | 57 | 160 | 719 | 0.3870 | 0.3870 | 0.3870 |
| 7 | 184 | 1017 | 5040 | 0.3082 | 0.3082 | 0.3082 |

TABLE IV
MINIMUM $SCC_{avg}$ VALUES FOR SHARING AN LFSR WITH 3 DIFFERENT SNGS BY EXTENDING THE CIRCULAR SHIFT APPROACH.

| $n$ | k1 | k2 | k3 | $SCC_{avg}$ $(C_{k1}, C_{k2})$ | $SCC_{avg}$ $(C_{k1}, C_{k3})$ | $SCC_{avg}$ $(C_{k2}, C_{k3})$ |
|---|---|---|---|---|---|---|
| 4 | 0 | 1 | 3 | 0.6254 | 0.6254 | 0.6254 |
| 5 | 0 | 3 | 4 | 0.4885 | 0.6507 | 0.6507 |
| 6 | 0 | 2 | 4 | 0.4626 | 0.4626 | 0.4626 |
| 7 | 0 | 2 | 5 | 0.4468 | 0.4468 | 0.3369 |
| 8 | 0 | 3 | 6 | 0.4373 | 0.2587 | 0.4373 |

TABLE V
MINIMUM $SCC_{avg}$ VALUES FOR SHARING AN LFSR WITH 3 DIFFERENT SNGS USING THE PROPOSED APPROACH WITH WBG.

| $n$ | P1 | P2 | P3 | $SCC_{avg}$ $(PL_{P1}, PL_{P2})$ | $SCC_{avg}$ $(PL_{P1}, PL_{P3})$ | $SCC_{avg}$ $(PL_{P2}, PL_{P3})$ |
|---|---|---|---|---|---|---|
| 4 | 3 | 10 | 23 | 0.5207 | 0.5207 | 0.5207 |
| 5 | 10 | 39 | 119 | 0.4321 | 0.4276 | 0.3994 |
| 6 | 40 | 177 | 720 | 0.3260 | 0.3260 | 0.3260 |
| 7 | 184 | 1017 | 5040 | 0.2381 | 0.2381 | 0.2381 |

TABLE VI
Actual runtimes (in *seconds*) of Algorithm 1 using the exact $SCC_{avg}$ values and the similarity function.

| method | $m$ | $n=4$ | $n=5$ | $n=6$ | $n=7$ |
|---|---|---|---|---|---|
| $SCC_{avg}$ | 3 | 0.021 | 1.089 | 162.247 | 37174.120 |
|  | 4 | 0.045 | 4.897 | 987.482 | 301105.142 |
| $S(k)$ | 3 | 0.012 | 0.060 | 2.112 | 225.098 |
|  | 4 | 0.026 | 0.217 | 14.006 | 1538.910 |

TABLE VII
Repeating the experiment of Table III using $S(k)$.

| $n$ | P1 | P2 | P3 | $SCC_{avg}$ $(PL_{P1}, PL_{P2})$ | $SCC_{avg}$ $(PL_{P1}, PL_{P3})$ | $SCC_{avg}$ $(PL_{P2}, PL_{P3})$ |
|---|---|---|---|---|---|---|
| 4 | 5 | 12 | 13 | 0.6071 | 0.5470 | 0.5207 |
| 5 | 23 | 46 | 61 | 0.4882 | 0.4887 | 0.4885 |
| 6 | 92 | 232 | 291 | 0.4052 | 0.4489 | 0.4119 |
| 7 | 597 | 1392 | 1729 | 0.3422 | 0.3351 | 0.3385 |

For more complex examples, we compare different implementations of 31- and 267-tap FIR filters in the form of a MUX tree explained in [6] and [7]. For both filters, we use MATLAB to generate low-pass filters' coefficients. For the SBoNG method, we use the SNG circuit described in [16] with 8-bit LFSRs, and for the circular shifting method, we use circuits similar to the architectures described in [6]. When a separate LFSR is used for each data and selection input, the number of required LFSRs for each application is listed in Table VIII. As the table shows, for the 267-tap filter, the proposed approach provides better accuracy compared to the circular shift method. For this filter, the no-share LFSR and SBoNG methods can achieve lower MSE values but their hardware complexity is more.

Finally, we apply our technique to the implementation of two image processing applications, i.e., edge detection and image segmentation, and compare their stochastic computation using different circuits. We evaluate our method by the Roberts cross edge detection algorithm implemented in SC [25] and by the kernel density estimation (KDE)-based image segmentation [3]. In the circuits related to the no-share LFSR method, each data and selection input of the multiplexers uses a separate LFSR. We obtained MSE values by exploiting five normalized grayscale still images with 256 levels from black to white for the edge detection algorithm, and four grayscale movies with 33 frames for the image segmentation algorithm. Table VIII lists MSE values for each algorithm calculated by taking the average of the MSE values of all trials for a design. For the edge detection, our method results in a MSE value close to that of the no-share LFSR and SBoNG method, however, with lower hardware complexity. For the KDE-based image segmentation, our proposed circuit leads to a MSE value nearly half of the MSE value for the circular shift circuit with the same hardware complexity.

TABLE VIII
CALCULATED MSE FOR DIFFERENT SC IMPLEMENTATIONS OF MULTIPLICATION, 31-TAP FIR FILTER, 267-TAP FIR FILTER, EDGE DETECTION, AND KDE-BASED IMAGE SEGMENTATION.

| Application | #of LFSR | No-share LFSR | Simple-share | SBoNG | Circular Shift | Proposed |
|---|---|---|---|---|---|---|
| Multiplier | 2 | 0.00008 | 0.01057 | 0.00010 | 0.00012 | 0.00001 |
| f31 | 61 | 0.00062 | 0.01825 | 0.00064 | 0.00065 | 0.00065 |
| f267 | 533 | 0.00814 | 0.04912 | 0.00883 | 0.00947 | 0.00902 |
| Edge detection | 7 | 0.00055 | 0.11714 | 0.00056 | 0.00052 | 0.00054 |
| KDE | 96 | 0.02285 | 0.88158 | 0.03582 | 0.08349 | 0.04324 |

## VII. CONCLUSION

In this paper, we investigated the design of low-cost and low-correlated SNG circuits using LFSR sharing. To reduce the correlation among the generated bit streams, we permuted the output of a shared LFSR before using it as input for different SNGs. We modeled the behavior of $SCC_{avg}$ for all permutations and our results show that for an LFSR's output, its first permutation in the reverse lexicographic order provides the minimum cross correlation. Compared to prior work with the same hardware complexity, i.e., the circular shift [8][6], our method results in SNGs with lower cross correlation values. We also proposed an algorithm for finding a set of $m$ permutations that can be shared among $m$ SNGs with minimum cross correlation. We used the proposed SNGs in the SC-based implementation of several applications and the results show that, with low hardware complexity, we obtain better computational accuracy compared to prior methods.

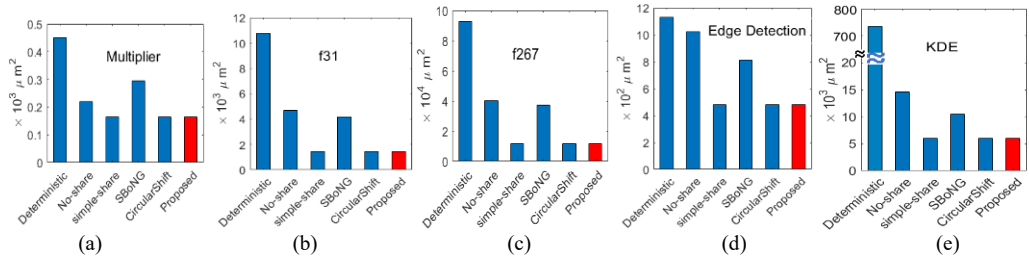

Fig7- Area comparison for different implementations of (a) multiplier, (b) 31-tap FIR filter, (c) 267-tap FIR filter, (d) edge detection, and (e) KDE-based image segmentation.